\documentclass[review, 1p]{elsarticle}

\usepackage{lineno,hyperref}
\modulolinenumbers[5]

\journal{Journal of \LaTeX\ Templates}

\usepackage{url}
\usepackage[utf8]{inputenc}
\usepackage{tabularx}

\usepackage{orcidlink}
\usepackage{graphicx}
\usepackage{dcolumn}
\usepackage{bm}
\usepackage[utf8]{inputenc}
\usepackage[T1]{fontenc}
\usepackage{mathptmx}
\usepackage{etoolbox}
\usepackage{xcolor}
\usepackage{graphicx}
\usepackage{lineno,hyperref}
\usepackage{blindtext}
\usepackage{epsfig}
\usepackage{latexsym}
\usepackage{graphics}
\usepackage{epsfig}
\usepackage{amsmath,amssymb,amsfonts,theorem,color,bm}
\usepackage{multirow, multicol, boldline}
\usepackage{stfloats} 
\usepackage{float}
\usepackage{subfig}
\usepackage{soul}
\usepackage{caption}
\hypersetup{colorlinks,allcolors=black}
\usepackage[normalem]{ulem}

\newcommand\etal{{\it et al. }}
\newcommand\ie{{\it i.e. }}

\begin{document}

\begin{frontmatter}

\title{Efficient Reduced Order Modeling Based on HODMD to Predict Intraventricular Flow Dynamics}

\newcommand{\orcidEL}{0000-0002-4514-6471}
\newcommand{\orcidJGM}{0000-0002-7422-5320}
\newcommand{\orcidSLCM}{0000-0003-3605-7351}

\author[addressUPM]{E. Lazpita \orcidlink{\orcidEL}}
\cortext[mycorrespondingauthor]{Corresponding author}
\ead{e.lazpita@upm.es}

\author[addressUPM,addressCSC]{J. Garicano-Mena \orcidlink{\orcidJGM}}
\author[addressUPM,addressCSC]{S. Le Clainche \orcidlink{\orcidSLCM}}

\address[addressUPM]{ETSI Aeron\'autica y del Espacio - Universidad Polit\'ecnica de Madrid, 28040 Madrid, Spain}
\address[addressCSC]{Center for Computational Simulation (CCS), 28660 Boadilla del Monte, Spain}

\begin{abstract}
Accurate and efficient modeling of cardiac blood flow is crucial for advancing data-driven tools in cardiovascular research and clinical applications. Recently, the accuracy and availability of computational fluid dynamics (CFD) methodologies for simulating intraventricular flow have increased. However, these methods remain complex and computationally costly. This study presents a reduced order model (ROM) based on higher order dynamic mode decomposition (HODMD). The proposed approach enables accurate reconstruction and long term prediction of left ventricle flow fields. The method is tested on two idealized ventricular geometries exhibiting distinct flow regimes to assess its robustness under different hemodynamic conditions. By leveraging a small number of training snapshots and focusing on the dominant periodic components representing the physics of the system, the HODMD-based model accurately reconstructs the flow field over entire cardiac cycles and provides reliable long-term predictions beyond the training window. The reconstruction and prediction errors remain below 5\% for the first geometry and below 10\% for the second, even when using as few as the first 3 cycles of simulated data, representing the transitory regime. Additionally, the approach reduces computational costs with a speed-up factor of at least $10^{5}$ compared to full-order simulations, enabling fast surrogate modeling of complex cardiac flows. These results highlight the potential of spectrally-constrained HODMD as a robust and interpretable ROM for simulating intraventricular hemodynamics. This approach shows promise for integration in real-time analysis and patient specific models.
\end{abstract}


\end{frontmatter}

\section{\label{sec:Introduction} Introduction}

Cardiovascular diseases remain the leading cause of mortality worldwide, with heart failure representing a major source of morbidity and a significant reduction in quality of life \cite{WHO2024}. In this context, understanding and predicting intraventricular blood flow dynamics is essential to improving diagnosis, prognosis, and treatment planning in patients with cardiac dysfunction (see Ref. \cite{Thomas2006Assessment, Kilner2000Asymmetric}). The left ventricle (LV), as the main pumping chamber of the heart, plays a central role in cardiac output, and its hemodynamics are intimately linked to both structural integrity and functional performance \cite{Pedrizzetti2015Left, Sengupta2006Left}.

Computational fluid dynamics (CFD) has emerged as a powerful tool for investigating intraventricular flows, complementing \textit{in vivo} measurements by enabling detailed, controllable exploration of flow patterns. Numerous studies have applied CFD to both idealized and patient-specific LV geometries, providing valuable insights into vortex dynamics, energy dissipation, and wall shear stresses. For instance, Spandan \etal \cite{Spandan2017Parallel} introduced a parallel interaction potential approach coupled with the immersed boundary method for fully resolved fluid--structure interaction simulations of deformable cardiac structures. This methodology has been extended to simulate healthy and pathological ventricles with natural and prosthetic mitral valves \cite{Meschini2018Flow}, as well as to investigate systolic anterior motion in hypertrophic cardiomyopathy \cite{Meschini2021Systolic}.

Domenichini \etal \cite{Domenichini2005Three} conducted early numerical studies on the three-dimensional inflow into a ventricular-shaped cavity. They later complemented this with experimental validation \cite{Domenichini2007Combined}, providing deeper insights into the mechanisms driving intraventricular flow. More recently, they explored the interaction between myocardial motion and fluid dynamics, demonstrating how alterations in tissue displacement are reflected in hemodynamic force distributions \cite{Domenichini2016Hemodynamic}. Quarteroni and collaborators have developed advanced frameworks for cardiac modeling that combine electromechanics, fluid dynamics, and patient-specific data \cite{Quarteroni2017Integrated, Africa2024Lifex}. These contributions reinforce the value of CFD as a predictive tool in cardiac research.

These works represent only a small sample of the growing body of literature. A comprehensive overview of CFD-based cardiac flow simulations, including their application to vortex ring (VR) analysis, is provided in the review by Mahesh \etal \cite{Nagargoje2025Review}. The VR forms during early diastolic filling and plays a critical role in the efficient reorientation of the blood influx towards the aortic valve throughout the cardiac cycle. Its formation and stability are closely tied to ventricular geometry, wall motion, and pathological changes, making it a relevant biomarker for cardiovascular health. In a recent study (see Ref. \cite{Lazpita2025Characterizing}), we analyzed the physical mechanisms underlying VR dynamics in two idealized LV geometries, identifying the VR as the dominant flow structure driving the system's temporal evolution and highlighting its potential as a diagnostic indicator.

Despite these advancements, generating high-fidelity CFD simulations of cardiac flows remains computationally demanding. The need to account for large deformations, moving boundaries, and physiologically realistic conditions results in considerable computational cost. Moreover, many CFD solvers suffer from numerical instability when applied to complex ventricular geometries or pathological flow regimes, often requiring extensive tuning or simplification of the physical model. These challenges restrict the scope of CFD applications in both clinical and research contexts.

As a result, many studies are constrained to a limited number of cardiac cycles, often fewer than five, impeding the statistical robustness and generalizability of the resulting data \cite{Kheradvar2012Vortex, Chnafa2014Image}. This limitation poses a significant barrier to the development of data-driven models, which require temporally rich datasets to achieve reliable performance across varying cardiac conditions.

In order to address these challenges, a novel reduced order model (ROM) is proposed. This model is designed to accelerate the obtention of intraventricular flow databases as an alternative to high-fidelity CFD simulations. Although ROMs are widely used in fluid mechanics to reduce computational cost, their application to cardiac hemodynamics is still limited. This is mainly due to the complexity of the problem, which involves large deformations, moving boundaries, strong nonlinearities and fluid-structure interaction \cite{Ballarin2016Fast, Lassila2014Model}. Our method extends the capabilities of existing modal decomposition frameworks by leveraging the higher order dynamic mode decomposition (HODMD) algorithm \cite{LeClainche2017Higher}. The HODMD technique extracts the dominant coherent structures of the flow and models their temporal evolution. In doing so, it captures the underlying physics, most notably the VR dynamics that govern early diastolic filling in the left ventricle.

Although HODMD has been successfully applied in simplified fluid scenarios, such as bluff body wakes or laminar shear flows \cite{LeClainche2017ROM}, this work presents its first application, to the authors' knowledge, to cardiac flows. In this context, we address additional challenges inherent to cardiac flow simulations, such as significant domain deformation driven by myocardial motion (\ie involving moving meshes). These complexities are further augmented by the influence of the ventricular wall on VR dynamics, as well as by the substantial temporal variability in velocity and pressure fields throughout the cardiac cycle, driven by the strong coupling between geometry deformation and flow evolution. By adapting the ROM to the spatiotemporal characteristics of the data, we achieve significant computational savings while preserving the essential dynamics of the system.

The following sections are structured as follows: Section~\ref{sec:Method} introduces the ROM methodology based on HODMD and its implementation. Section~\ref{sec:FlowDynamics} presents the two idealized left ventricle databases used in this study. In Section~\ref{sec:Results}, we provide a detailed analysis of the ROM configuration using two distinct idealized LV geometries. Finally, Section~\ref{sec:Conclusions} summarizes the main findings and discusses future directions.

\section{\label{sec:Method} Reduced Order Modelling}

As mentioned, the ROM employed in this study is based on HODMD, a data-driven technique that extracts dominant spatiotemporal patterns from complex dynamical systems. To handle the inherently high-dimensional nature of the simulation data, we adopt a tensor-based extension of the algorithm, known as multi-dimensional higher order dynamic mode decomposition (mdHODMD) \cite{Hetherington2024Modelflows}. This formulation allows us to preserve the multi-dimensional structure of the dataset and enhance both computational efficiency and interpretability.

The input data consists of time-resolved three-dimensional velocity fields obtained from CFD simulations. For each time instant, the velocity field is represented by three components defined over a structured grid in three spatial directions. By stacking all temporal snapshots together, the dataset can be naturally arranged as a five-dimensional tensor \( \boldsymbol{V} \in \mathbb{R}^{J_1 \times J_2 \times J_3 \times J_4 \times K} \), where:
\begin{itemize}
    \item \( J_1 = 3 \) corresponds to the velocity components \( (v_x, v_y, v_z) \),
    \item \( J_2, J_3, J_4 \) represent the spatial discretization density along the three coordinate directions, and
    \item \( K \) denotes the number of time steps.
\end{itemize}

The mdHODMD algorithm is then applied to this tensor in three consecutive main steps: dimensionality reduction via tensor decomposition, extraction of dynamic modes and their associated frequencies, and prediction of the system’s temporal evolution.

\subsection{\label{subsec:dimensionality} Data Dimensionality Reduction}
The processing of large-scale datasets directly poses a significant computational challenge, requiring the implementation of dimensionality reduction techniques to ensure manageability. Singular value decomposition (SVD)~\cite{Sirovich1987Turbulence} is widely used in fluid dynamics to reduce data dimensionality and filter out noise. However, the intraventricular flow simulations considered in this work are high-dimensional, transient, and highly non-linear. They involve moving geometries subject to large temporal deformations and complex spatiotemporal patterns that appear. As a result, more advanced techniques are required to analyze and predict their behavior effectively. In this study, we apply the high order singular value decomposition (HOSVD)~\cite{LeClainche2017Higher}, a tensor-based extension of SVD that enables simultaneous decomposition along multiple data dimensions. This method improves compression and noise filtering by decoupling the modes in each dimension. As a result, it can distinguish and eliminate noise or redundancies independently across spatial, temporal, and physical-magnitude components. Such enhanced denoising and compression capabilities make HOSVD particularly well-suited for building robust, low-rank representations from transient flow data \cite{Groun2025Eigenhearts}, thereby supporting the construction of accurate and efficient predictive models.

The HOSVD is applied to the 5D tensor $\mathbf{V}$, which in turn is factorized as
\begin{equation}
  \mathbf{V}_{j_1j_2j_3j_4k} \simeq \sum_{p_1=1}^{P_1}\sum_{p_2=1}^{P_2}\sum_{p_3=1}^{P_3}\sum_{p_4=1}^{P_4} \sum_{n=1}^{N} \mathbf{S}_{p_1p_2p_3p_4n} \mathbf{W}^{(v)}_{j_1p_1}\mathbf{W}^{(x)}_{j_2p_2} \mathbf{W}^{(y)}_{j_3p_3}\mathbf{W}^{(z)}_{j_4p_4}\mathbf{T}_{kn},
  \label{eq:HOSVD}
\end{equation}
where $\mathbf{S}_{p_1p_2p_3p_4n}$ is the 5-th dimensional core tensor, and the orthonormal matrices $\mathbf{W}^{(v)}$, $\mathbf{W}^{(x)}$, $\mathbf{W}^{(y)}$, $\mathbf{W}^{(z)}$, and $\mathbf{T}$ contain the SVD modes for each dimension. A tunable tolerance \( \varepsilon \) determines the number of retained modes \( N \), ensuring that \( \sigma_{N +1}^{(j)}/\sigma_1^{(j)} \leq \varepsilon_1 \) with \(j = 1,\ldots,5\). This temporal matrix $\mathbf{T}$ containing the temporal coefficients associated with each mode (of dimensions $N \times K$) serves as input for the subsequent dynamic mode decomposition (DMD) computation.

\subsection{\label{subsec:HODMD} Calculation of DMD Modes}
DMD~\cite{Schmid2010Dynamic} provides a reduced-order modeling approach by expressing complex spatiotemporal data as a linear combination of \( M \) coherent structures, or modes, that evolve in time. The approximation of the dataset \( \mathbf{V}_{j_1}(x,y,z,t_k) \) (where \( j_1=1,\ldots,3 \) indexes the velocity components) is formulated as
\begin{equation}
    \mathbf{V}_{j_1}(x,y,z,t_k) \approx \sum_{m=1}^{M} a_m \, \mathbf{u}_m(x,y,z) \, e^{(\delta_m + i\omega_m)t_k} \quad \text{for} \; k=1,\ldots,K \, ,
    \label{eq:DMDexpansion}
\end{equation}
where \( \mathbf{u}_m \) denotes the spatial structure of each DMD mode, \( a_m \) is the mode amplitude, \( \delta_m \) is the temporal growth rate, and \( \omega_m \) represents the oscillation frequency. 

To obtain these modes, DMD assumes a linear mapping between consecutive data snapshots \( \{ \mathbf{v}_1, \mathbf{v}_2, \dots, \mathbf{v}_K \} \), posing the relation to be described as
\begin{equation}
    \mathbf{V}_2^{K} \approx \mathbf{R} \, \mathbf{V}_1^{K-1} \, ,
    \label{eq:Koopman}
\end{equation}
where \( \mathbf{R} \) is a linear operator that approximates the underlying dynamics, effectively capturing the temporal progression through the so-called Koopman framework \cite{Mezic2013Analysis, Otto2021Koopman}.

HODMD \cite{LeClainche2017Higher} extends standard DMD by incorporating multiple time-delayed snapshots, enhancing robustness and accuracy in capturing complex flow structures. Furthermore, the mdHODMD algorithm preserves the tensorial formulation by applying the higher-order Koopman assumption to the temporal matrix $\mathbf{T}$ extracted in the previous step using HOSVD (see Eq.~\ref{eq:HOSVD}) as
\begin{equation}
    \mathbf{T}_{d+1}^K \approx \widehat{\mathbf{R}}_1 \mathbf{T}_1^{K-d}+ \widehat{\mathbf{R}}_2 \mathbf{T}_2^{K-d+1} + \ldots + \widehat{\mathbf{R}}_d \mathbf{T}_d^{K-1}.
    \label{eq:HigherOrderKoopman}
\end{equation}

This formulation links each snapshot to $d$ previous time steps, capturing intricate temporal dynamics beyond what standard DMD offers. The modified Koopman matrix $\widehat{\mathbf{R}}$, which contains all the linear operators (\( \hat{R_1}, \hat{R_2}, \ldots, \hat{R_d} \)), is then computed to extract the DMD modes. Another tunable tolerance is introduced for the HODMD; in this work, and following the guidelines discussed in Ref. \cite{LeClainche2017Higher, Hetherington2024Modelflows} which is set equal to the one selected for the HOSVD, such that \( \varepsilon_1 = \varepsilon_2 = \varepsilon \).

\subsection{\label{subsec:prediction} Prediction}

Once the data is decomposed into DMD modes, future-state predictions are obtained by evaluating Eq.~\eqref{eq:DMDexpansion} at time instants beyond the training window \( t > t_K \). Specifically, the model is constructed using data within the interval \( t \in [0,\, pT] \), where \( T \) is the period of the cardiac cycle and \( p \) is the number of cycles employed for training. In this work, we will consider \( p \in \mathbb{R} \) and \(p\leq 10 \). The resulting reduced-order model is then used to predict the flow dynamics in a later interval, namely \( t \in [10T,\, 20T] \), allowing us to assess the long-term predictive capabilities of the method.

The way modes are selected plays a crucial role in ensuring prediction accuracy and stability in the model. Depending on the calibration of the algorithm, spurious modes unrelated to the underlying physics may be captured, typically associated with noise or numerical artifacts. As a result, using all computed \( M \) DMD modes can degrade the quality of the extrapolated prediction. To address this difficulty, we adopt a filtering strategy in which only those modes with growth rates lower than a certain threshold are deemed physically relevant and retained for prediction. This threshold is set with a tunable growth rate value \( \delta_{tune} \). Thus, stable modes will correspond to \( -\delta_{tune} < \delta_m<0 \) with \(\delta_{tune}>0\). Almost all of the selected modes correspond to frequencies associated with the dominant dynamics of the system.

Furthermore, when working with transient simulation data, certain modes, although physically relevant, may exhibit highly negative growth rates, leading to their rapid attenuation over time. This offers the potential to enhance the temporal stability of the model predictions; we also construct an alternative formulation in which we enforce \( \delta_m = 0 \), thereby eliminating exponential decay. This modification prevents excessive damping of key flow structures and enables more reliable long-term forecasts of the intraventricular dynamics.

To assess the accuracy of the ROM, we analyze the absolute error distribution over the entire dataset by computing the histogram of the normalized absolute error, defined as:
\begin{equation}
    E(\%) = \frac{|\textbf{v} - \textbf{v}_{\text{pred}}|}{\max |\textbf{v} - \textbf{v}_{\text{pred}}|} \cdot 100 \, ,
    \label{eq:AbsoluteError}
\end{equation}
where $\textbf{v}$ denotes the reference simulation data and $\textbf{v}_{\text{pred}}$ is the reconstructed prediction. The normalization factor corresponds to the maximum absolute error computed independently for each velocity component, considering all spatial locations and time steps in the domain.

The number of bins in the histogram is determined using Sturges's formula, applied to the spatial grid:  
\begin{equation}
    N_{\text{bins}} = 1 + \log_2(N_{\text{grid}}) \, ,
\end{equation}
where $N_{\text{grid}} = N_x \times N_y \times N_z$ represents the total number of spatial points. In our case, this results in approximately 20 bins, allowing for an intuitive interpretation: the first bin represents the relative frequency of errors within the range 0--5\%, the second within 5--10\%, and so forth.

A key advantage of the ROM is its drastic reduction in computational cost. In this study, this increase in computational efficiency is quantified using the speed-up factor (SUF), defined as:  
\begin{equation}
    \text{SUF} = \frac{t_{CFD} \times N_{CPU}}{t_{ROM}}
\end{equation}
where $t_{\text{CFD}}$ is the runtime of the full-order CFD simulation, $N_{\text{CPU}}$ is the number of processors used, and $t_{\text{ROM}}$ is the time required for the ROM prediction.

\section{\label{sec:FlowDynamics} Flow Dynamics}

To evaluate the performance of the ROM introduced earlier, we apply it to two distinct datasets, both of them representing idealized left ventricle geometries. The first model, referred to as \texttt{Ideal 1}, consists of a semi-ellipsoidal chamber extracted from Ref.~\cite{Zheng2012Computational} and will be used to develop and illustrate the methodology. The second model, \texttt{Ideal 2}, has a smoother, more rounded chamber shape and has been studied in prior works Ref.~\cite{Vedula2014Computational, Kjeldsberg2023Verified}. This model will serve as an additional test case to evaluate the robustness of the approach under different conditions.

\subsection{\label{subsec:Database} Database}

To ensure that the idealized geometries represent physiologically realistic left ventricular conditions, particularly the end-systolic volume (ESV) that corresponds to the minimum volume at the end of systole, we define a set of non-dimensional geometric parameters. These parameters are expressed in Tab. \ref{tab:geometry_parameters} using the base radius \( a \) of the ventricular chamber as the reference length for each model.

\begin{table}[H]
\centering
\begin{tabular}{| l | c | c |}
\hline
\textbf{Parameter} & \texttt{Ideal 1} \cite{Zheng2012Computational} & \texttt{Ideal 2} \cite{Vedula2014Computational, Kjeldsberg2023Verified} \\
\hline
Chamber length ratio \( b/a \)   & 4.0 & 2.2 \\
Inlet diameter \( D/a \)         & 1.2 & 0.85 \\
Outlet diameter \( d/a \)        & 0.4 & 0.6 \\
Tube height \( H/a \)            & 2.4 & 2.7 \\
Inlet offset \( C/a \)           & 0.275 & 0.35 \\
Outlet offset \( c/a \)          & 0.675 & 0.6 \\
Reference radius \( a \) [cm]    & 2.0 & 2.9 \\
\hline
\end{tabular}
\caption{Non-dimensional geometric parameters for the two idealized ventricular models, using base radius \( a \) as reference length.}
\label{tab:geometry_parameters}
\end{table}

\begin{figure}[h]
    \centering
    \includegraphics[trim = 0 0 100 0, clip, width=\textwidth]{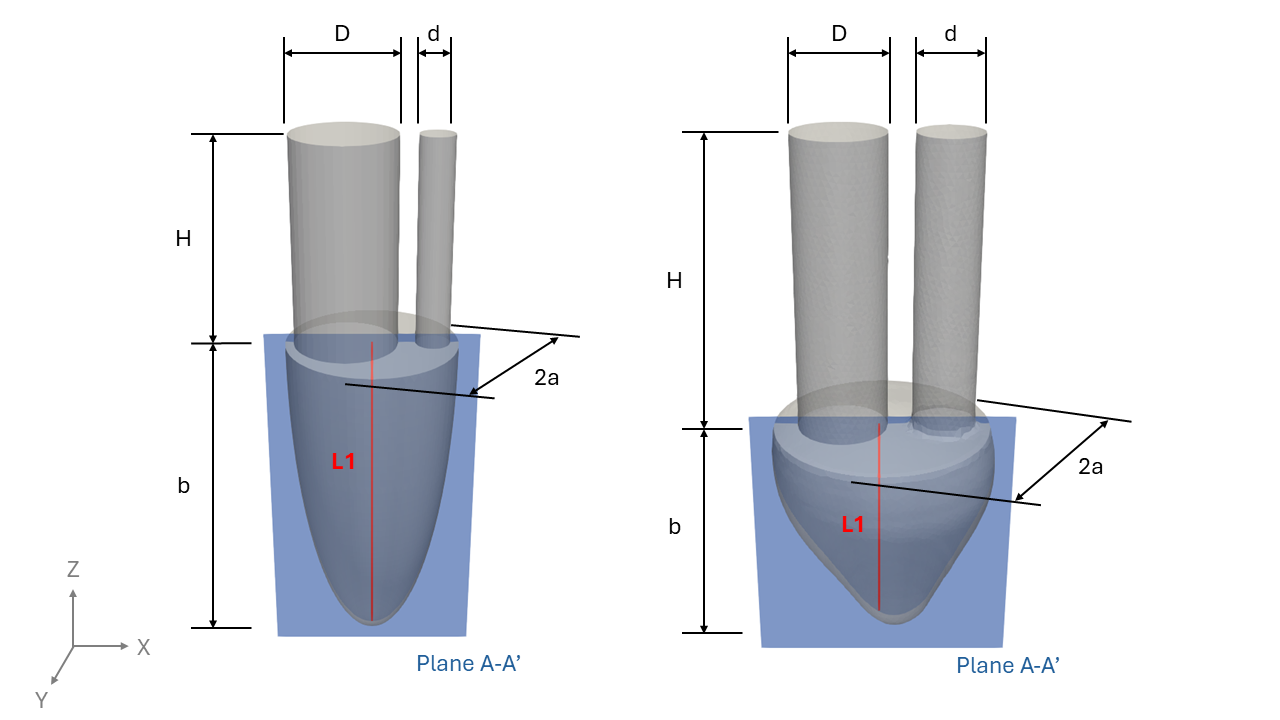}
    \caption{Representation of both idealized LV models: (left) \texttt{Ideal 1} and (right) \texttt{Ideal 2}. Blue: Symmetry plane A-A' used for two-dimensional visualization. Red: Probe line L1 to represent temporal data.}
    \label{fig:geometries}
\end{figure}

A schematic representation of these parameters and their spatial arrangement is provided in Fig.~\ref{fig:geometries}, illustrating the geometry and orientation of each feature. The figure also identifies the symmetry plane A-A' (located at \(y=0\)) used throughout the analysis (highlighted in blue). Additionally, a probe line (L1), indicated in red, is defined to extract temporal data at specific spatial locations. For the \texttt{Ideal 1} model, L1 is located at \( x/a = 0, \, y/a =  0 \), and it extends through all the vertical \(z\) of the chamber. For the \texttt{Ideal 2} model, L1 is positioned at \( x/a = 0, \, y/a = 0 \), and again it takes all the chamber height.

Using the two idealized geometries of the LV, we conducted CFD simulations to investigate intraventricular flow dynamics. Simulations were carried out using the solver \textit{Ansys Fluent}~\cite{Fluent}. In both cases, blood was modeled as an incompressible Newtonian fluid with constant density \(\rho = 1060~\text{kg/m}^3\) and dynamic viscosity \(\mu = 0.004~\text{Pa}\cdot\text{s}\).

To characterize the flow regime, we computed the Reynolds number (\(\mathrm{Re}\)) based on the inlet tube diameter \(D\) (non-dimensionalized as \(D/a\)) and the peak inlet velocity \(V_{\text{max}}\) over the cardiac cycle. The resulting Reynolds numbers were approximately \(\mathrm{Re} \approx 5500\) for the \texttt{Ideal 1} model and \(\mathrm{Re} \approx 4900\) for \texttt{Ideal 2}. Despite these values indicating transitional or potentially turbulent flow, we assumed a laminar regime, in line with previous studies~\cite{He2022Numerical, Tagliabue2017Complex}, where coherent vortex structures, such as the main VR, were accurately captured under laminar assumptions.

A zero velocity initial condition is set in the simulation pipeline for both models. Moreover, for both simulations, pressure-based boundary conditions were set at the inlet and outlet. These boundaries dynamically alternated between open and closed states to represent the cardiac cycle: during diastole, the inlet was open and the outlet treated as a wall; during systole, the roles were reversed. This boundary-switching was complemented by the dynamic motion of the ventricular wall, imposed by prescribing an interpolated sequence of meshes corresponding to each time instant. This approach induces pressure gradients within the chamber that drive blood in and out, thereby mimicking physiological behavior.

The simulation framework and its assumptions were extensively validated in previous studies~\cite{Lazpita2024ECCOMAS, Lazpita2024Modeling, Lazpita2025Characterizing}, including convergence analyzes and comparisons against benchmark data. All numerical simulations were performed on the high-performance computing (HPC) cluster \textit{Magerit}, hosted at CeSViMa (Centro de Supercomputación y Visualización de Madrid). The computations were parallelized using 40 MPI tasks within a single node.

For the rest of the manuscript, time is non-dimensionalized using the period \( T \), such that \( t^* = t/T \). This normalization allows for a consistent comparison of corresponding time instants across both simulation databases. It is also important to note that, for the first geometry, the diastolic phase takes place within the interval \( t^* \in [i,\, i + 0.67] \), followed by the systolic phase in \( t^* \in [i + 0.67,\, i + 1] \). In contrast, for the second geometry, diastole spans \( t^* \in [i,\, i + 0.8] \), and systole occurs within \( t^* \in [i + 0.8,\, i + 1] \), where \( i \) denotes the cycle index.

A key aspect of intraventricular hemodynamics is the formation and evolution of a vortex ring during diastole, triggered by the inflow through the mitral valve \cite{Nagargoje2025Review}. This structure undergoes tilting and eventual breakdown due to instabilities, and its dynamics are considered a primary indicator of pumping efficiency. Accurately capturing the vortex ring behavior is, therefore, essential.

\begin{figure}[h]
    \centering
    \includegraphics[trim = 50 0 50 0, clip, width=\textwidth]{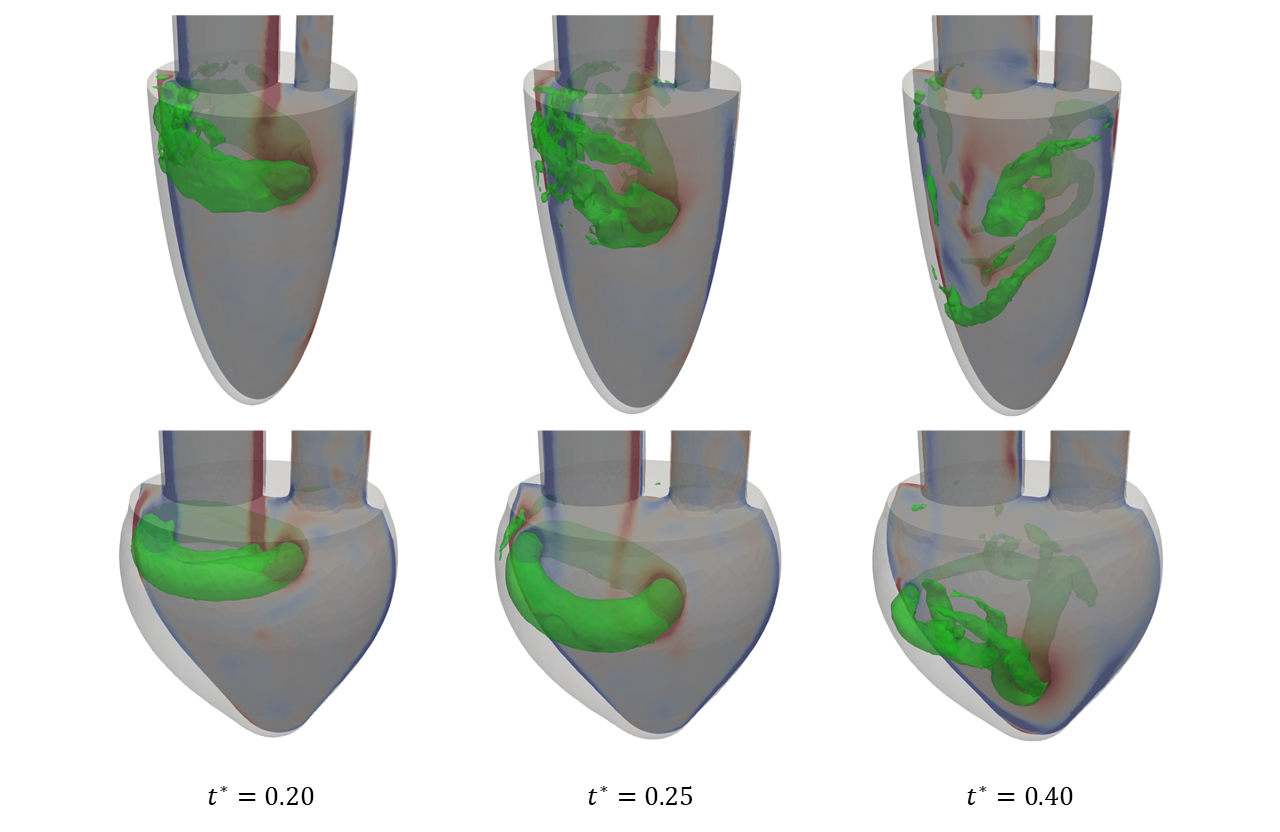}
    \caption{Representative snapshots of the intraventricular flow fields for \texttt{Ideal 1} (top) and \texttt{Ideal 2} (bottom). Green: Q-Criterion isosurface at 2500. Background: vorticity in the symmetry plane, range $[-100, 100]$.}
    \label{fig:database}
\end{figure}

Figure~\ref{fig:database} illustrates representative time instants for both geometries. The green isocontours show the Q-Criterion at a level of 2500, highlighting vortex cores, while the colormap on the symmetry plane displays the out-of-plane vorticity in the range \([-100, 100] \: \text{s}^{-1}\). Notably, the vortex ring in \texttt{Ideal 1} dissipates earlier than in \texttt{Ideal 2}; by \( t^* = 0.25 \), the ring in \texttt{Ideal 1} is already breaking up, whereas \texttt{Ideal 2} still maintains a coherent and smooth structure. Even at \( t^* = 0.40 \), the second model, though deformed, maintains a closed vortex ring, while the first model’s ring has fully dissipated. 

As shown in Lazpita \etal \cite{Lazpita2025Characterizing}, the vortex breakdown process is strongly influenced by ventricular geometry. In the \texttt{Ideal 1} configuration, which features a narrower chamber, the vortex interacts with the ventricular wall at an earlier stage. This early interaction triggers the breakdown process before the vortex ring can reach the apex. In contrast, the \texttt{Ideal 2} model, with a wider chamber, allows the vortex ring to fully develop, travel toward the apex, and break down as a combination of a reduction in velocity and wall interaction.

These contrasting behaviors can be attributed to two main factors:
\begin{itemize}
    \item The narrower geometry of \texttt{Ideal 1} causes the vortex to interact more strongly with the ventricular walls, preventing it from reaching the apex and accelerating its breakdown
    \item \texttt{Ideal 2} exhibits a lower ejection fraction and volume change, resulting in reduced pressure gradients and slower flow dynamics.
\end{itemize}

These differences in vortex ring dynamics will be key for assessing the ROM's ability to capture flow physics across varying conditions.

Each simulation database consists of 20 cardiac cycles, with flow fields sampled at a time interval of \( \Delta t^* = 0.05 \): therefore, these are 20 snapshots per cycle and a total of \( K = 400 \) snapshots. For training the ROM, we use data from the first 10 cycles, which include the initial transient phase. The remaining 10 cycles are reserved for validation, as they capture the dynamics once the system has reached temporal convergence. This distinction is further supported by the application of DMD, which reveals that growth rates stabilize only in the latter cycles.

Although it is common practice in the literature to simulate up to 6 cycles, we shall find that the flow remains in a transient state at that point \cite{Cane2022CFD, Grunwald2022Intraventricular, Korte2023Hemodynamic}. This highlights the importance of longer simulations to ensure the reliability of the ROM and the physical relevance of the extracted modes.

The cases listed in Tab.~\ref{tab:cases} are analogous for both LV geometries and thus we maintain the same denomination for consistency and brevity. Each case corresponds to a specific time window and number of cycles, which are used either for training or validation of the ROM. We also report the tunable growth rate value selected for each case, which enables the identification of persistent modes that represent the main dynamics of the system, effectively discarding the spurious modes extracted by HODMD.

\begin{table}[H]
\centering
\begin{tabular}{| l | c | c | c | c | c |}
\hline
\textbf{Case} & \(p\) & \textbf{Time Interval \( t^* \)} & \textbf{Snapshots \(K\)} & \textbf{Purpose} & \( \delta_{tune} \) \\
\hline
T-1.5 & 1.5  & [0, 1.5]    & 30   & Training (transient)   & 1.0  \\
T-3   & 3    & [0, 3]      & 60   & Training (transient)   & 0.5  \\
T-10  & 10   & [0, 10]     & 200  & Training (transient)   & 0.01 \\
V-10  & 10   & [10, 20]    & 200  & Validation (converged) & 0.01 \\
\hline
\end{tabular}
\caption{Summary of simulation cases used to build and evaluate the ROM discussed in Sec.~\ref{sec:Method}, including the number of cardiac cycles \(p\), normalized time intervals \( t^* \), snapshot count \(K\), their intended purpose, and the selected tunable growth rate parameter \( \delta_{tune} \).}
\label{tab:cases}
\end{table}

\subsection{\label{subsec:Uncertainty} Database Sensitivity Assessment}

Despite careful simulation setup, the complex nature of moving-wall CFD simulations introduces inherent variability across cycles, even setting the boundary conditions perfectly periodic \cite{Chnafa2014Image}. While major flow structures, such as the primary vortex ring, secondary vortices, and outflow jets, are consistently reproduced, smaller-scale features might vary.

We quantify the variability across cardiac cycles by analyzing the cycle-to-cycle variation of the total kinetic energy (TKE), defined at each non-dimensional time instant \( t^* \) as:
\begin{equation}
    \mathrm{TKE}(t^*) = \int_{V(t^*)} \frac{1}{2}\rho \mathbf{v}^2\, dV,
\end{equation}
where \( V(t^*) \) is the time-dependent ventricular volume, \( \rho \) is the fluid density, and \( \mathbf{v} \) is the velocity vector. The total kinetic energy integrated over each cardiac cycle is computed as:
\begin{equation}
    \mathrm{TKE}_i = \int_{i-1}^{i} \mathrm{TKE}(t^*)\, dt^*, \quad i = 1, \ldots, 20,
\end{equation}
where \( i \), again, denotes the index of the cycle. To measure convergence and variability, we evaluate the relative root mean square error (RRMSE) of each cycle with respect to the last one:
\begin{equation}
    \mathrm{RRMSE}_{\mathrm{TKE}}(\%) = \frac{|\mathrm{TKE}_{20} - \mathrm{TKE}_i|}{|\mathrm{TKE}_{20}|} \cdot 100,
\end{equation}
where \( \mathrm{TKE}_{20} \) is the integrated kinetic energy for the 20th cycle, considered as the reference.

\begin{figure}[h]
    \centering
    \includegraphics[width=\textwidth]{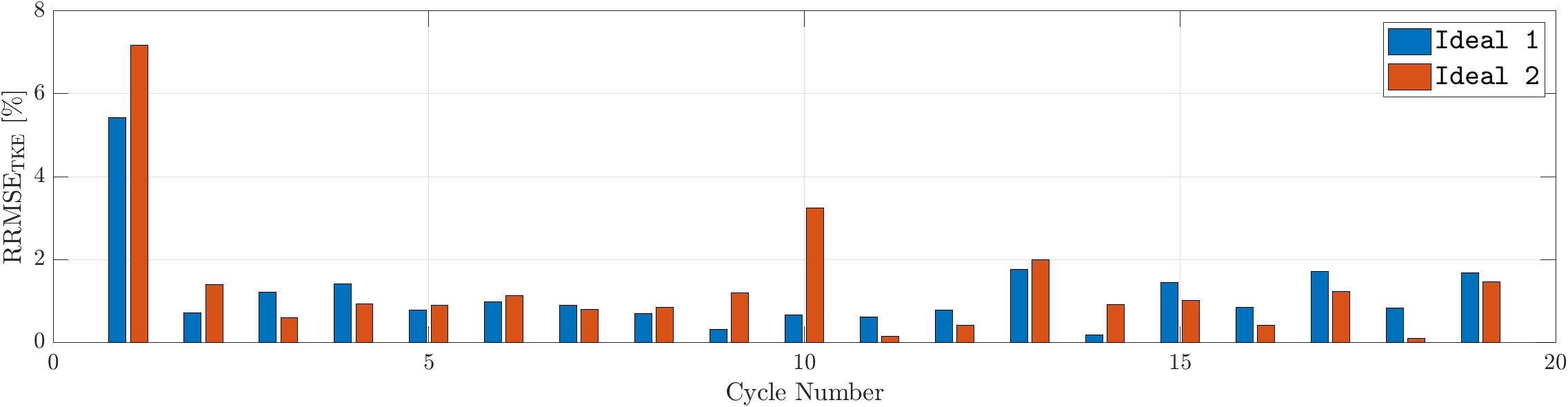}
    \caption{Cycle-to-cycle RRMSE of total kinetic energy for \texttt{Ideal 1} and \texttt{Ideal 2}, visualized using a grouped bar plot.}
    \label{fig:uncertainty_bar}
\end{figure}

Figure~\ref{fig:uncertainty_bar} illustrates the cycle-to-cycle variability in total kinetic energy error, expressed as the RRMSE for both \texttt{Ideal 1} and \texttt{Ideal 2} geometries. Remember that the simulation was initialized from a zero velocity condition. After the initial transient, the RRMSE remains below 4\% across all subsequent cycles, indicating good reproducibility of global flow features. However, local flow characteristics, such as instantaneous velocity fields, may still exhibit larger discrepancies due to intrinsic variability in the numerical solution~\cite{Chnafa2014Image, Cane2022CFD}. As such, predictions obtained from reduced-order models should be interpreted within a baseline uncertainty margin of approximately 5--10\%.

\subsection{\label{subsec:FrequencyAnalysis} True Frequency Analysis}

\begin{figure}[h]
    \centering
    \includegraphics[width=\textwidth]{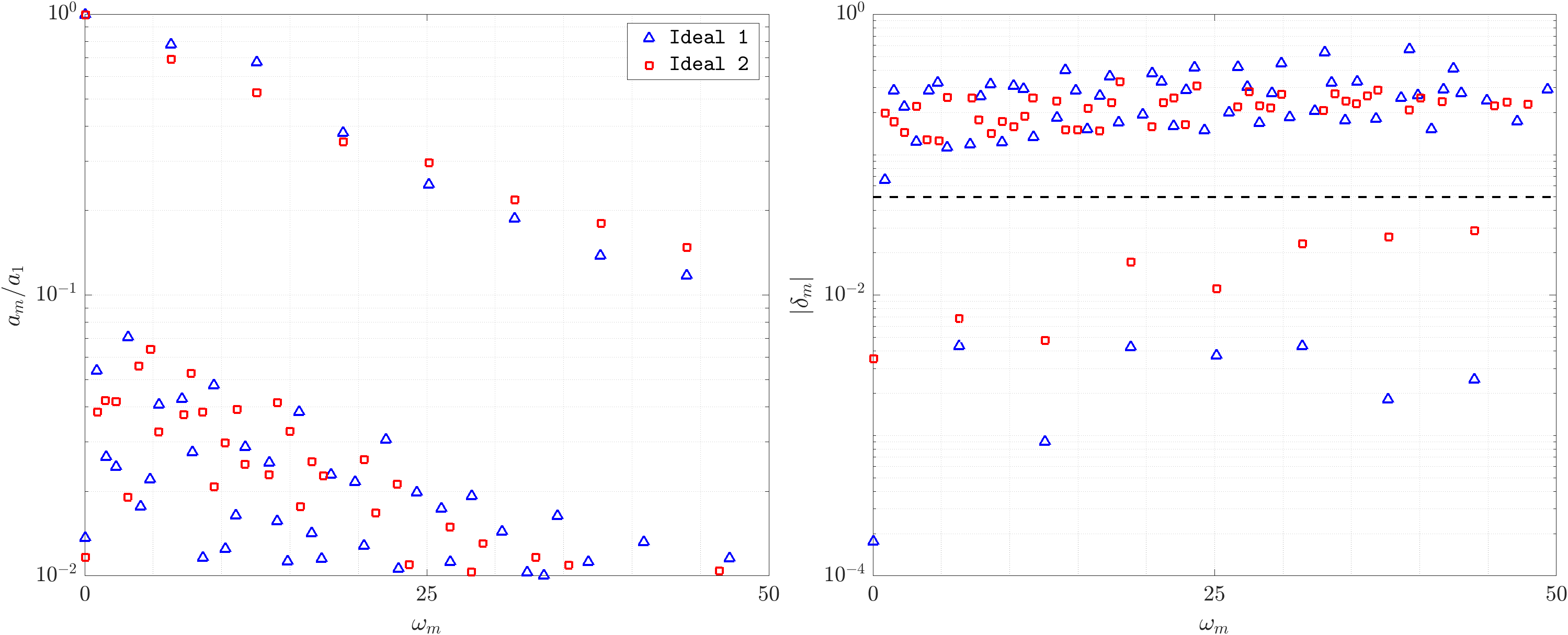}
    \caption{Spectrum of DMD modes for both idealized models: normalized amplitude vs. frequency (left) and absolute value of growth rate vs. frequency (right). The dashed line at \( \delta_{tune} = 5 \cdot 10^{-2}\) in the right panel separates persistent physical modes from transient or spurious ones.}
    \label{fig:DMD-spectrum}
\end{figure}

Figure~\ref{fig:DMD-spectrum} presents the DMD spectra obtained from a representative HODMD calibration applied to both \texttt{Ideal 1} and \texttt{Ideal 2} cases. The left panel shows the normalized amplitude of each mode with respect to frequency, while the right panel displays the corresponding absolute values of the growth rates. Although only one calibration is depicted here for clarity, the modes shown have been previously identified as robust through an exhaustive sensitivity analysis involving the HODMD parameters \(d\) and \(\varepsilon\), following the methodology described in Refs.~\cite{LeClainche2020Coherent, Hetherington2024Modelflows}.

The amplitude spectra in the left panel reveal the dominant frequencies that appear consistently across both geometries. These include the fundamental frequency \(\omega = 2\pi\) and its harmonics, such as \(4\pi\), \(6\pi\), and up to approximately \(14\pi\), indicating strong periodicity and physical relevance of these modes. Lower-amplitude modes, scattered throughout the spectrum, are not persistent across calibrations and are interpreted as numerical noise or transient artifacts.

The right panel provides additional insight through the analysis of growth rates \(|\delta_m|\). Robust physical modes tend to have small or nearly zero growth rates, indicating long-term persistence throughout the simulation. A threshold of \( \delta_{\text{tune}} = 5 \cdot 10^{-2} \) (shown as a dashed line) distinguishes these physical modes from the spurious ones, which exhibit larger decay or amplification rates and lack consistency. This threshold was established based on a qualitative shift observed in the spectral structure during the calibration study: below this value, the modes were consistently identified and physically meaningful, while above it, the spectrum became irregular and inconsistent.

These results confirm that the dominant dynamics of both \texttt{Ideal 1} and \texttt{Ideal 2} models are governed by a limited number of coherent structures oscillating at fundamental and harmonic frequencies, in agreement with the known periodicity of the cardiac cycle.

\section{\label{sec:Results} Results}

\subsection{\label{subsec:Ideal_1} Semi-ellipsoidal Model \texttt{Ideal 1}}
For the first model, HODMD is applied to analyze several training subsets, each consisting of a varying number of cardiac cycles \(p\) as defined in Table~\ref{tab:cases}. The objective is to determine how many cycles are necessary to accurately capture the temporal dynamics of the system. To evaluate this, we construct the temporal evolution of the system using the frequencies and growth rates extracted using HODMD. We consider two scenarios: (i) using the growth rates computed by HODMD, and (ii) setting all growth rates to zero.

\begin{figure}[h]
    \centering
    \includegraphics[width=\textwidth]{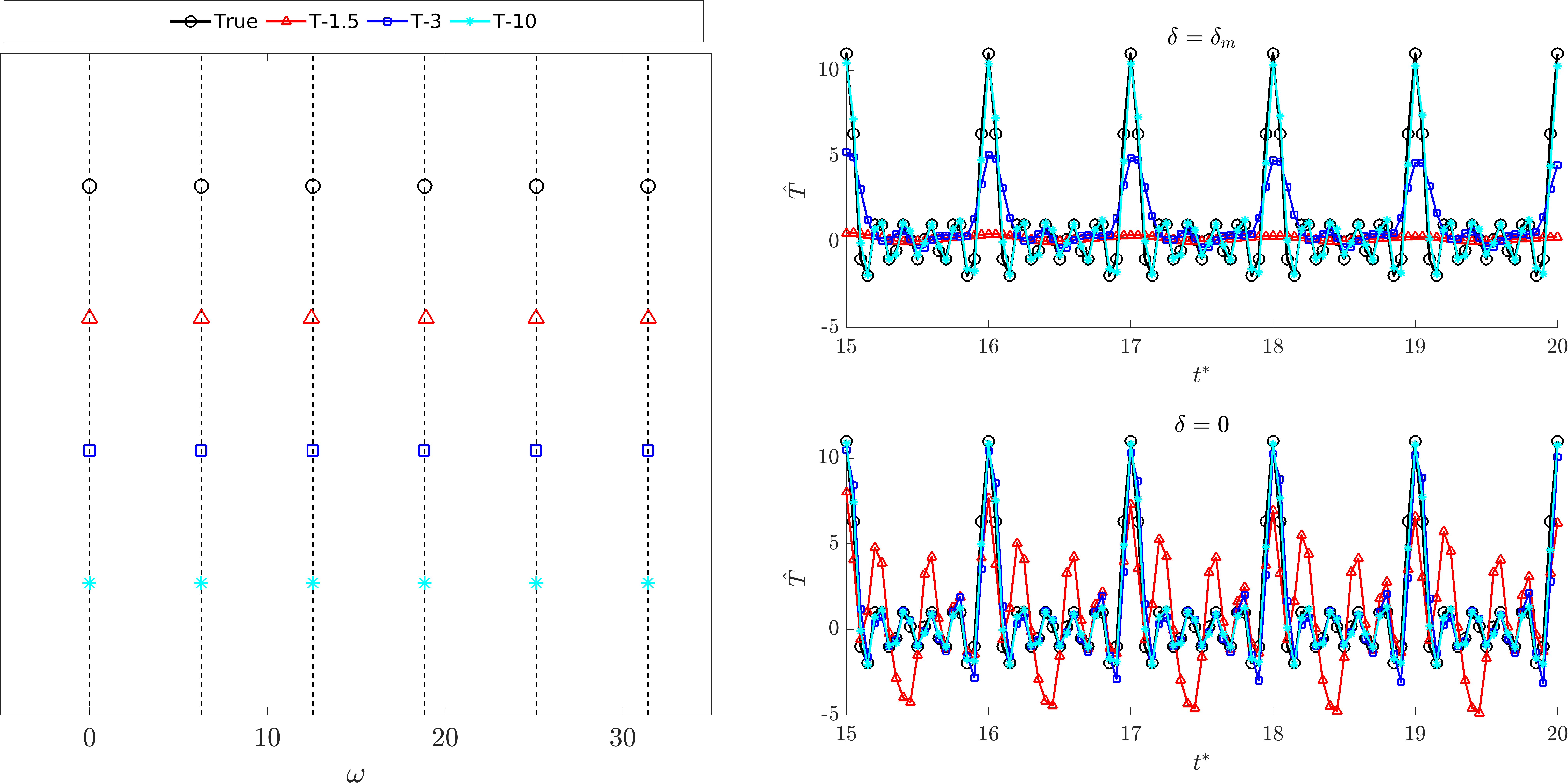}
    \caption{Comparison of the calculated frequency (left) and temporal coefficient (right) with HODMD against the true system (black) for the \texttt{Ideal 1} model: (right-top) computed growth-rate and (right-bottom) growth-rate set to zero. The cases are coloured as: (red) T-1.5, (blue) T-3, and (cyan) T-10 cases (see Tab. \ref{tab:cases}).}
    \label{fig:ideal1-error-freq}
\end{figure}

Figure~\ref{fig:ideal1-error-freq} provides a detailed evaluation of HODMD performance across different temporal window lengths. The left panel shows the estimated frequencies compared to the theoretical ones. The results confirm that, for all cases, the dominant frequency and its harmonics are accurately captured, with minimal deviation from the true system values.

The right panel illustrates the reconstructed temporal coefficients $\hat{T}$ for each case, comparing the original reconstruction that preserves the computed growth rates (top) against a modified version where all growth rates are manually set to zero (bottom), thereby enforcing purely periodic behavior.

In the shortest window (T-1.5, red), the dominant frequency is already reasonably well estimated; however, the temporal coefficients show significant distortion, primarily due to an excessive negative growth rate and the inaccurate capture of higher harmonics. Suppressing growth in this case provides some improvement but remains suboptimal. For the intermediate case (T-3, blue), frequency estimates are much more precise, and enforcing zero growth yields a nearly-periodic signal with minimal distortion. Finally, in the longest window (T-10, cyan), both frequency and growth rate estimates are sufficiently accurate, and the temporal evolution remains consistent with the physical oscillatory behavior even without modifications, leaving the zero-growth correction redundant.

The results indicate that the estimated growth rate for some modes is significantly negative (e.g., \( \delta \approx -0.3 \)), which causes an artificial attenuation of the signal over time with respect to the ground truth. Therefore, a negative growth rate leads to exponential decay, reducing the amplitude of the reconstructed signal and compromising long-term prediction accuracy. Setting the growth rate to zero significantly improves prediction stability by removing the artificial damping and better representing the periodic nature of these simulations.

Furthermore, including all modes introduces noise and spurious contributions, resulting in degraded accuracy, whereas the principal modes alone sufficiently capture the main dynamics. While 1.5 cycles are sufficient to recover the overall behavior, at least 3 cycles are needed for more precise predictions.

Now that the spectral characteristics of the simulations have been shown to be accurately captured using only three cycles (when the growth rate is set to zero), we turn our attention towards the spatial support. To this end, we perform predictions using DMD modes extracted from datasets of different lengths. The analysis focuses on the most accurate configurations, which are constructed by selecting modes with a growth rate below a defined threshold and subsequently enforcing a zero growth rate. This procedure ensures that all retained modes contribute equally over time.

\begin{figure}[h]
    \centering
    \includegraphics[width=\textwidth]{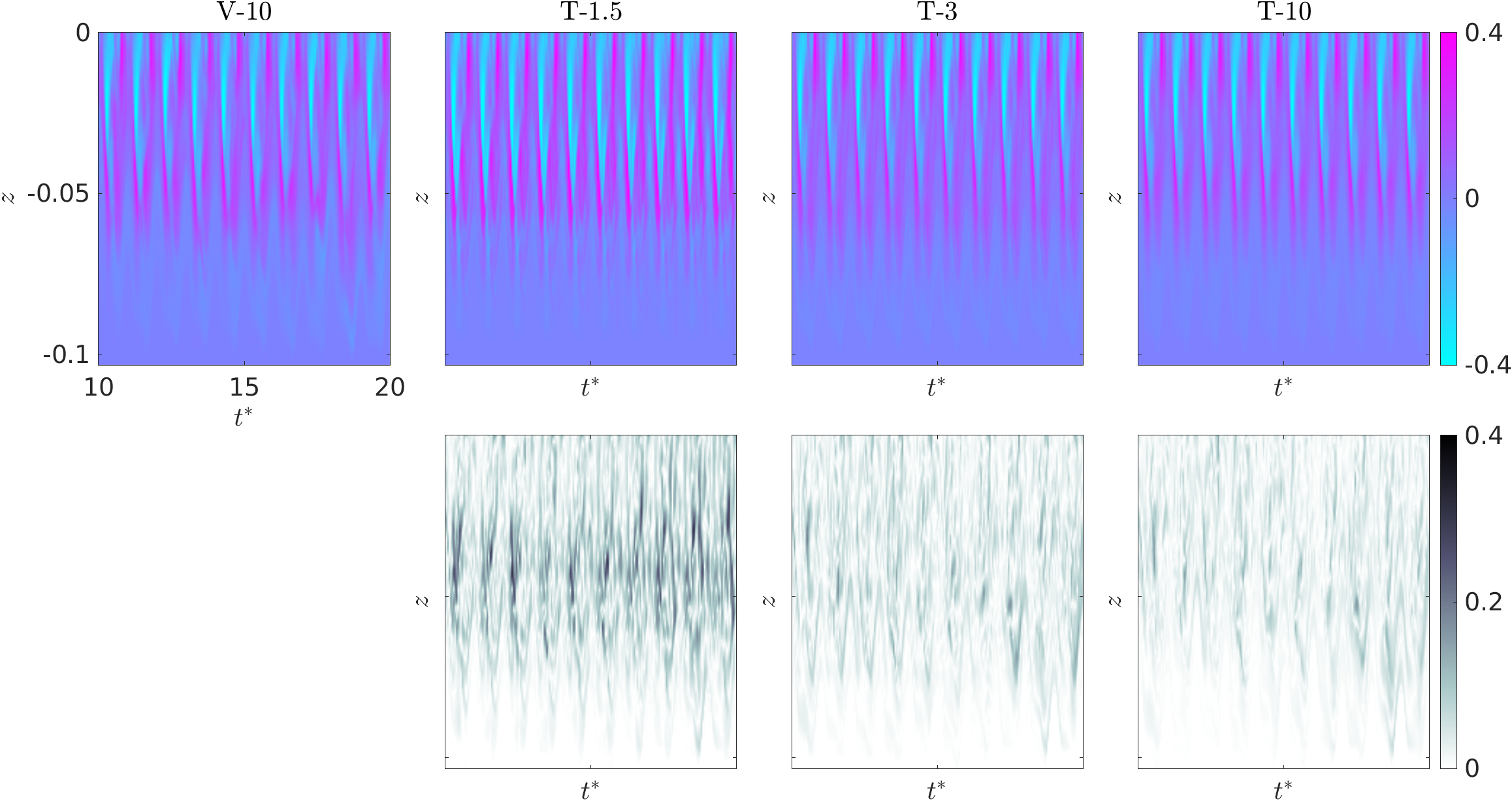}
    \caption{Temporal evolution of the \( v_x \) velocity component along center line L1 for the \texttt{Ideal 1} case. The top row shows predictions from reduced-order models constructed using training sets T-1.5, T-3, and T-10. The bottom row presents the absolute error relative to the reference case V-10. All results are obtained using a subset of selected modes with an enforced zero growth rate (\( \delta = 0 \)).}
    \label{fig:ideal1-error-line}
\end{figure}

Figure~\ref{fig:ideal1-error-line} illustrates the temporal evolution of the \( v_x \) velocity component along probe line L1. This component is particularly relevant because the flow exhibits a strong recirculating behavior around the vortex structure within the chamber. Consequently, \( v_x \) captures the dominant dynamics. Predictions obtained from different training sets are compared against the reference case V-10. The top row of the figure presents the predicted evolution, while the bottom row shows the corresponding absolute error with respect to the reference data.

The flow evolution at line L1 is influenced by both the inflow near the inlet and the recirculation induced by the vortex ring moving towards the apex. Across all configurations, accurate predictions are obtained by selecting only the permanent, physically relevant modes. This choice effectively removes spurious contributions that would otherwise corrupt the predictions. Enforcing a zero growth rate further improves accuracy by preventing any mode from dominating or decaying over time.

The results again show that using only 1.5 cycles of data leads to poor predictions with limited potential for improvement. Increasing the training length to three cycles significantly enhances accuracy, especially when the growth rate is set to zero, which prevents amplitude decay. Note that, even with ten cycles, small discrepancies persist, particularly near the inlet where the inflow dominates. This reflects the intrinsic difficulty of reconstructing the complex nonlinear behavior in that region. The overall results, despite these deviations, are consistent with the expected simulation accuracy and are considered reliable.
 
\begin{figure}[h]
    \centering
    \includegraphics[width=\textwidth]{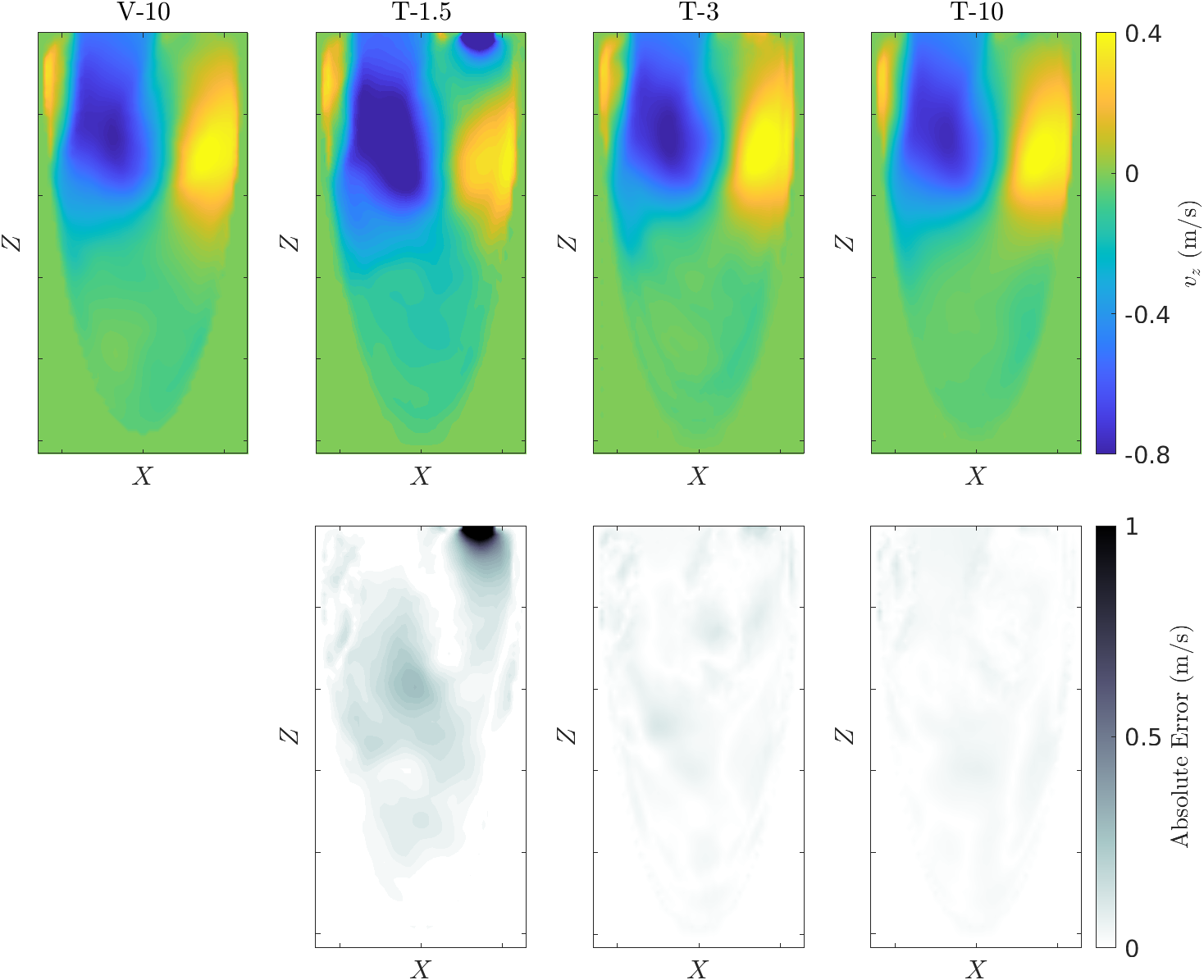}
    \caption{Comparison of the \(v_z\) velocity component in the A-A' plane at \(t^* = 19.25\) for the \texttt{Ideal 1} case using ROMs constructed from training sets T-1.5, T-3, and T-10 (top row), along with the corresponding absolute error with respect to the reference case V-10 (bottom row). The results displayed are using a selected subset of modes and a zero growth rate ($\delta = 0$) enforced.}
    \label{fig:ideal1-error-contour}
\end{figure}

After evaluating the accuracy of the line evolution over time, we performed a similar analysis to evaluate quality in a larger spatial region. Figure~\ref{fig:ideal1-error-contour} shows the two-dimensional contours of the \(v_z\) velocity component in the A--A' plane at \(t^* = 19.25\) (\(1/4\) of cycle) for the \texttt{Ideal 1} case, comparing the T-1.5, T-3, and T-10 configurations (top row). When compared to the reference case V-10, all configurations capture the main features of the flow field, including the inflow and recirculation zones within the chamber at this time instant. However, the T-1.5 case exhibits noticeable inaccuracies, particularly a nonphysical influence of the outflow that should not be present at this phase of the cycle.

The corresponding absolute error plots (bottom row) support the previous statement, computed as the absolute difference with respect to the V-10 solution. The T-1.5 configuration shows higher overall discrepancies, especially in the outlet region. In contrast, the T-3 case significantly reduces the error, achieving accuracy levels comparable to those of the T-10 configuration, which includes many more samples.

\begin{figure}[h]
    \centering
    \includegraphics[width=\textwidth]{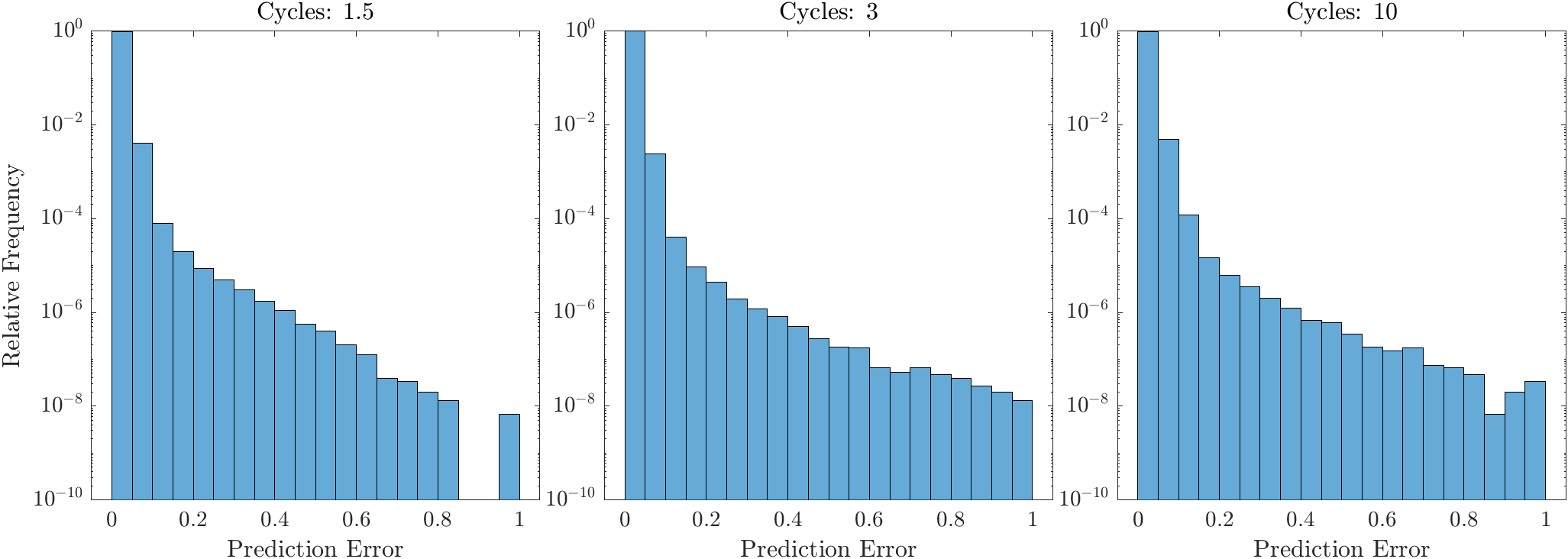}
    \caption{Histogram of the absolute error $E$ using 20 bins for the \texttt{Ideal 1} model (each bin corresponds to 5\% of error).}
    \label{fig:ideal1-error-histogram}
\end{figure}

A more quantitative evaluation of the error is shown in Fig.~\ref{fig:ideal1-error-histogram}. Here, the error histograms for this configuration over different numbers of cycles for, again, the most accurate configurations are presented. As observed, across all cases, in at least 99\% of the spatial locations, the error is below 5\%, demonstrating the reliability of the ROM.

It is important to acknowledge the intrinsic variability present in these types of simulations. Even when comparing different cycles of the ground truth simulation, minor discrepancies arise. Given this inherent variability, the accuracy achieved by the ROM is particularly noticeable, considering its significantly lower computational cost, which will be discussed shortly.

The greatest potential of using the ROM lies in its significant reduction of computational cost. For instance, the full-order CFD simulation was executed using 40 CPUs and required approximately 38 hours to compute 10 cardiac cycles with a well-resolved mesh configuration. In contrast, the ROM prediction over the same time span was completed in just 36 seconds, yielding a speed-up factor (SUF) of \(1.5 \times 10^5\). 

However, a trade-off must be established between computational savings and reconstruction accuracy. While shorter training windows significantly reduce the offline cost, as observed in case T-1.5, the ROM derived from such an economic (but poor) dataset may fail to capture essential flow features, leading to poor performance. In contrast, our analysis shows that using 3 cycles (case T-3) attains a reasonable balance: it provides sufficient temporal information to capture the dominant dynamics with acceptable accuracy, while still offering notable reductions in computational effort compared to longer simulations such as T-10.

\subsection{\label{subsec:Ideal_2} Experimentally-Derived Model: \texttt{Ideal 2}}

Having analyzed the tuning and performance of the ROM with the first idealized geometry, we now extend our study to the \texttt{Ideal 2} model. This serves as a test case to evaluate the method’s performance on a different geometry, where the intraventricular flow dynamics are notably different and evolve at a slower pace. More details about the flow patterns and the development of flow instabilities in this geometry can be found in \cite{Lazpita2025Characterizing}.

The frequency accuracy obtained with HODMD is comparable to the \texttt{Ideal 1} case shown in Fig.~\ref{fig:ideal1-error-freq}; thus, a detailed analysis is omitted for brevity. It is important to recall, however, that although the dominant frequency was accurately captured using only 1.5 cycles, the associated growth rate was not. As a result, improving the temporal predictions required either increasing the number of cycles or enforcing a zero growth rate. As before, the following results are obtained using a subset of selected modes with an enforced zero growth rate (\( \delta = 0 \)).

\begin{figure}[h]
    \centering
    \includegraphics[width=\textwidth]{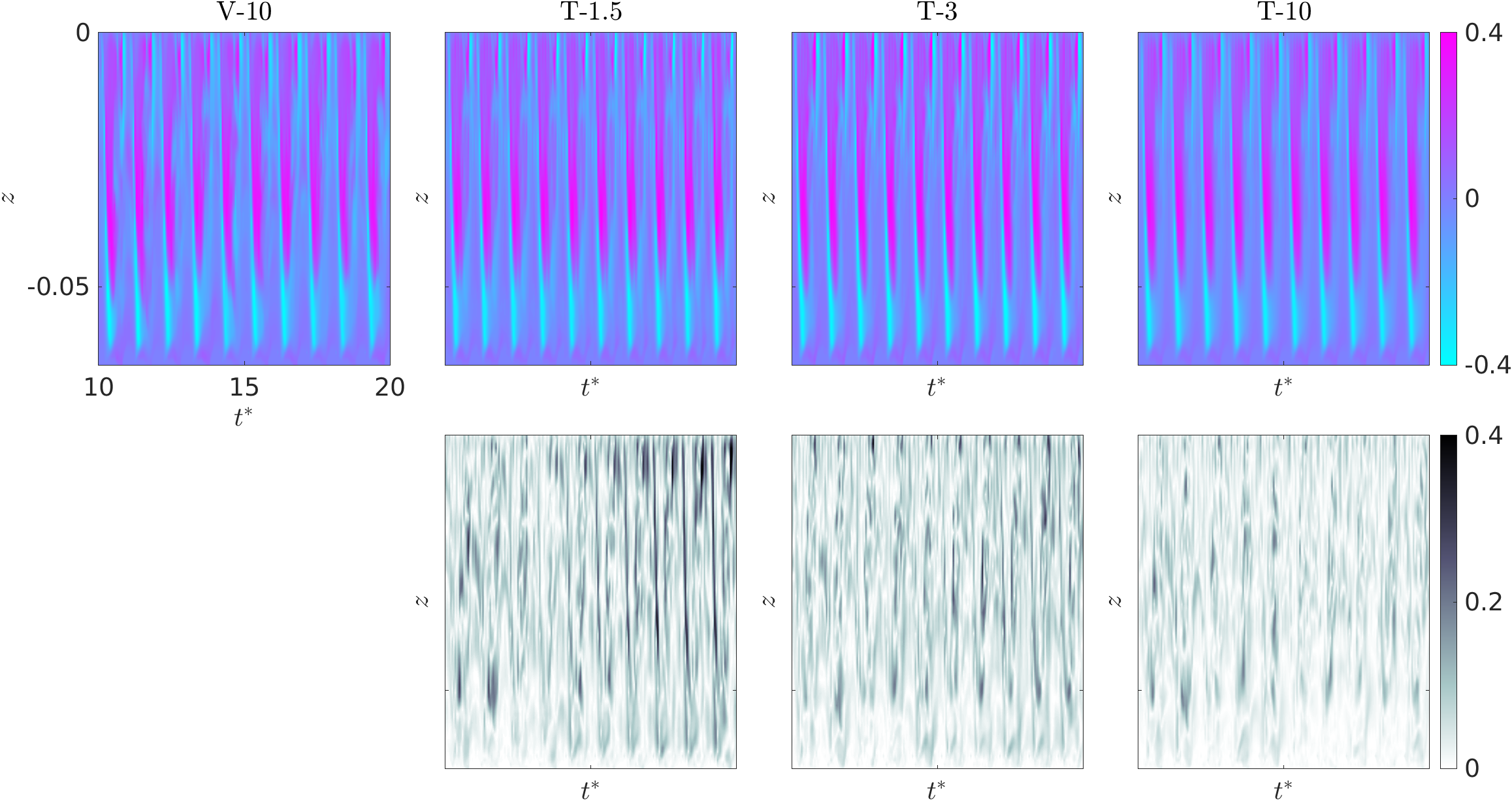}
    \caption{Counterpart of Fig.~\ref{fig:ideal1-error-line} for the \texttt{Ideal 2} model.}
    \label{fig:ideal2-error-line}
\end{figure}

A central part of the analysis focuses on the evolution of the \( v_x \) velocity component along the center line (L1) within the chamber. This evaluation compares ROM predictions with the original CFD data. Figure~\ref{fig:ideal2-error-line} presents the temporal evolution of \( v_x \) at L1 using a selected subset of modes enforcing zero growth rate (\( \delta = 0 \)). The absolute error between the ROM predictions and the reference data is also shown.

In this case, the flow exhibits slower dynamics due to smaller volume variations when compared with the \texttt{Ideal 1} configuration \cite{Lazpita2025Characterizing}. This adds new challenges to the prediction task. The model trained with only 1.5 cycles shows the highest error, although the error levels are comparatively smaller than for the previous case. Increasing the training set length to three cycles improves the predictions, yet the accuracy remains limited and comparable to that achieved with ten cycles.

\begin{figure}[h]
    \centering
    \includegraphics[width=\textwidth]{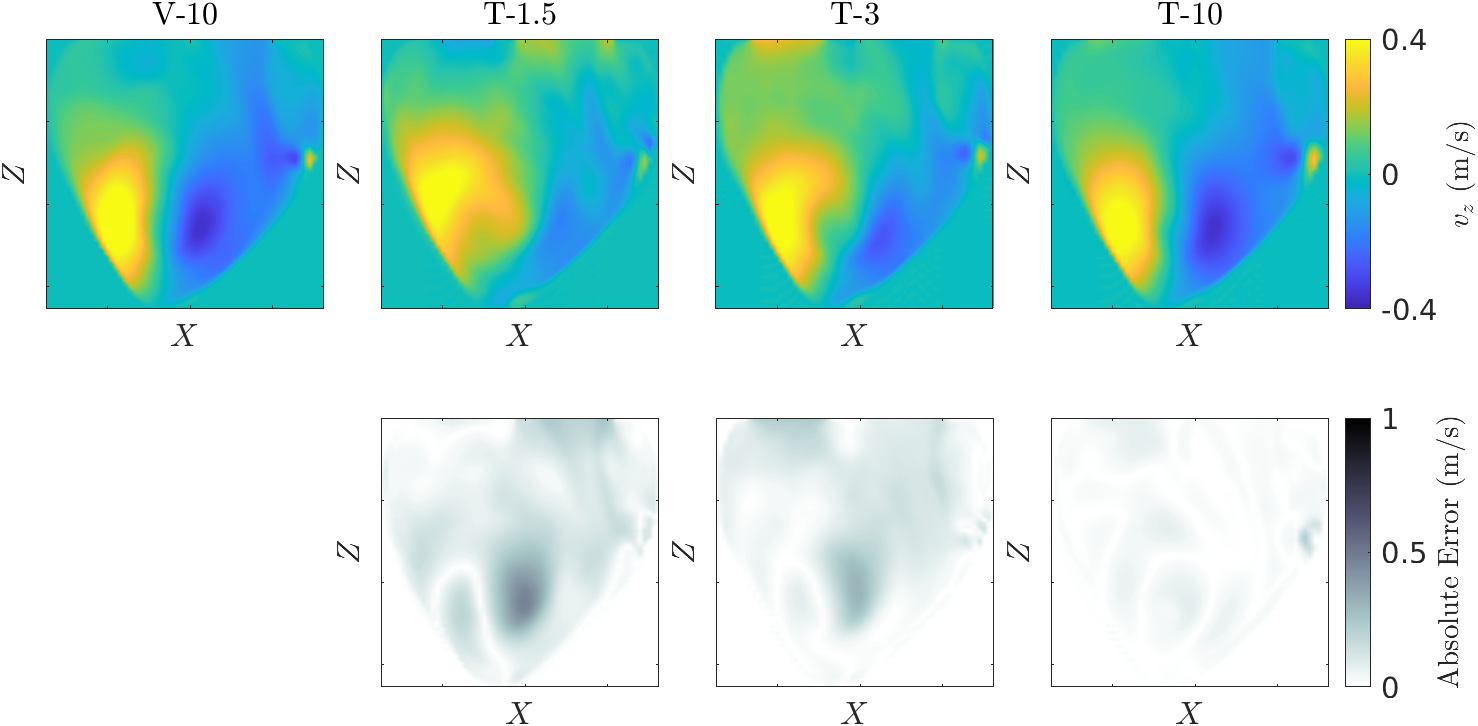}
    \caption{Counterpart of Fig.~\ref{fig:ideal1-error-contour} for the \texttt{Ideal 2} model.}
    \label{fig:ideal2-error-contour}
\end{figure}

Again, spatial resolution is assessed in Fig.~\ref{fig:ideal2-error-contour}, showing the contours of the \(v_z\) velocity component in the A-A' plane at \(t^* = 19.25\) (\(1/4\) of cycle) for the configurations yielding the most accurate results for cases T-1.5, T-3, and T-10, along with the associated absolute error computed against the reference case V-10 for the \texttt{Ideal 2} case. The analysis of the two-dimensional contours in this geometry reveals that, for the T-1.5 case, the error is already quite low, although some localized inaccuracies remain. Nevertheless, the general flow features are well captured. Accuracy improves notably when the ROM is constructed using three cycles (T-3), and is significantly enhanced for the T-10 case, where the training data is richer. Despite this, the results obtained with T-3, or even T-1.5, are particularly appealing given the substantial computational savings achieved when only 1.5 or 3 cycles are needed instead of 10 or more (see later).

Comparing these results with the ones obtained in \texttt{Ideal 1} Sec.~\ref{subsec:Ideal_1}, the results for 1.5 cycles have improved, while T-3 or T-10 both offer good results for either model. This could be linked to the simpler vortex evolution that the second model has compared to the complex early breakdown seen in the first one (see Ref.~\cite{Lazpita2025Characterizing}), which makes the dynamics easier to capture.

\begin{figure}[h]
    \centering
    \includegraphics[width=\textwidth]{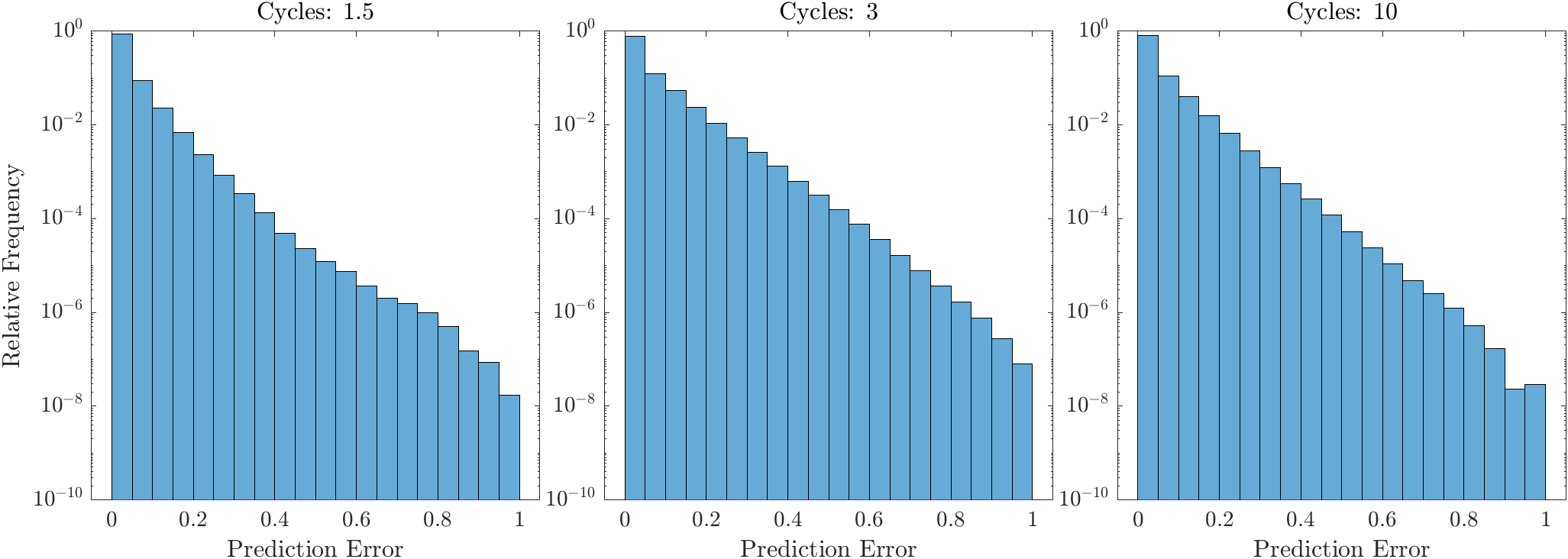}
    \caption{Counterpart of Fig.~\ref{fig:ideal1-error-histogram} for the \texttt{Ideal 2} model.}
    \label{fig:ideal2-error-histogram}
\end{figure}

As before, we use error histograms to assess the relative frequency of different error magnitudes. Figure~\ref{fig:ideal2-error-histogram} shows the distribution of the normalized absolute error across predictions. The relative frequency of errors below 5\% is close to 90\%, indicating accurate overall performance, even though the overall error is slightly higher than in \texttt{Ideal 1}.

These results are even better when considering the computational cost reduction, as reflected in the SUF. In this idealized model, the high-fidelity CFD simulation required 48 hours on 40 CPUs to compute the cycles between \( t=10 \) and \( t=20 \). In contrast, the ROM executed the same cycles in just 41 seconds, yielding an SUF of \( 1.7 \times 10^{5} \).

As already discussed at the end of Sec.~\ref{subsec:Ideal_1}, there is a trade-off between ROM accuracy and computational cost. For this second model, training with 1.5 or 3 cycles greatly reduces data and offline cost compared to the full 10 cycles, while maintaining acceptable accuracy even for T-1.5. Nonetheless, T-3 should be preferred for improved performance.

\section{\label{sec:Conclusions} Conclusions}
This work presented a predictive reduced-order model (ROM) based on higher-order dynamic mode decomposition (HODMD) to accelerate the acquisition of data on intraventricular flows in idealized left ventricular (LV) geometries, as compared to computational fluid dynamics (CFD) simulations. Unlike traditional uses focused on data reconstruction or analysis (\cite{Lazpita2022Generation, Corrochano2024Hierarchical, Lazpita2024Data}, the ROM here proposed is specifically designed to predict flow evolution efficiently. This end is attained by constructing a modal expansion and adjusting the temporal evolution of the modes to forecast the dynamics in regions of interest. This methodology, previously applied in fluid mechanics problems for accelerating CFD simulations~\cite{LeClainche2017ROM, LeClainche2017Higher}, is here employed for the first time, to the authors' knowledge, in the context of cardiac flows, where the complexity increases due to moving wall boundary conditions, nonlinear vortex dynamics, and confined, time-varying geometries.

The ROM was applied to two representative cases with distinct geometries and flow behavior: one characterized by early vortex breakdown (\texttt{Ideal 1}) and another dominated by a more stable vortex ring (\texttt{Ideal 2}). To ensure accurate predictions, a robust modal expansion is required; hence, the HODMD framework was selected \cite{Hetherington2024Modelflows}. This method enabled the extraction of dominant modes and frequencies from limited training data, mitigating the effects of noise and non-permanent components. Different calibrations of the method were tested, including the enforcement of zero growth rates to stabilize long-term predictions.

The results demonstrated that HODMD can effectively identify and reproduce the key flow structures, such as the formation, development, and eventual breakdown of the main vortex ring, and their associated frequencies with high accuracy. When combined with spectral filtering and growth rate tuning, the model achieved improved interpretability and reduced sensitivity to spurious components.

Quantitatively, the ROM achieved low relative errors and high correlation with the reference CFD results in both time and space, even when trained with a small number of cycles. Furthermore, the computational gains were substantial: while full-order simulations required dozens of hours on multi-core architectures, the ROM predictions were executed in seconds on a single core of a laptop computer, yielding speed-up factors (SUFs) of \(O(10^5)\). Reducing the training data to just 1.5 or 3 cycles instead of 10 further enhances the model’s efficiency. However, this also introduces a trade-off: using fewer cycles further lowers computational cost but may compromise accuracy. In our study, a training length of 3 cycles emerged as a suitable compromise between performance and efficiency.

It is also important to emphasize that training or simulating only 3 or 6 cycles, as done in many previous studies, is not sufficient to capture the full flow dynamics. These shorter intervals often reflect transient behavior. The present study clearly demonstrates the need for longer time series to achieve reliable and physically meaningful predictions.

In summary, this HODMD-based ROM combines high accuracy, physical interpretability, and improved computational efficiency, positioning it as a promising tool for accelerating cardiac flow simulations. Its applicability to fast parametric studies and real-time evaluations makes it especially relevant for future clinical scenarios. Ongoing and future work will focus on extending this framework to patient-specific geometries, integrating clinical imaging data, and exploring its capabilities in pathological conditions.

\section{Data Availability Statement}
The data that support the findings of this study are available upon reasonable request.
\\
Should the reader wish to gain a more detailed understanding of the process by which the databases were obtained, as well as other pertinent information, they are invited to visit the following website, where they will find the relevant codes and tutorials in Ref. \cite{ModelFLOWsCardiac}.

\section{Acknowledgements}
The authors acknowledge the grants TED2021- 129774B-C21 and PLEC2022-009235 funded by MCIN/AEI/ 10.13039/501100011033 and by the European Union “NextGenerationEU”/PRTR and the grant PID2023-147790OB-I00 funded by MCIU/AEI/10.13039/501100011033/FEDER, UE. The authors gratefully acknowledge the Universidad Politécnica de Madrid (www.upm.es) for providing computing resources on Magerit Supercomputer.

The authors gratefully acknowledge Prof. Vedula for kindly providing the geometry data of \texttt{Ideal 2} model that was essential for conducting this study.

\bibliography{biblio}

\end{document}